\documentclass[sigconf]{acmart}
\AtBeginDocument{%
  \providecommand\BibTeX{{%
    \normalfont B\kern-0.5em{\scshape i\kern-0.25em b}\kern-0.8em\TeX}}}

\acmConference[Tech Report]{Tech Report}{March 2021}{Stanford, CA}
\acmBooktitle{Technical Report}

\setcopyright{none}
\renewcommand\footnotetextcopyrightpermission[1]{}
\setcopyright{none}
\makeatletter
\renewcommand\@formatdoi[1]{\ignorespaces}
\makeatother

\usepackage{float}
\usepackage{pdfpages}

\begin{document}

\title{Crowdsourcing County-Level Data on Early COVID-19 Policy Interventions in the United States: Technical Report}

\author{Jacob Ritchie}
\email{jritchie@stanford.edu}
\affiliation{%
  \institution{Stanford University}
  \city{Stanford}
  \state{California}
  \country{USA}
}

\author{Mark Whiting}
\email{markew@seas.upenn.edu}
\affiliation{%
  \institution{University of Pennsylvania}
  \city{Philadelphia}
    \city{Pennsylvania}
  \country{USA}}

\author{Sorathan Chaturapruek}
\email{sorathan@cs.stanford.edu}
\affiliation{%
 \institution{Stanford University}
 \city{Stanford}
 \state{California}
 \country{USA}}

\author{J.D. Zamfirescu-Pereira}
\email{zamfi@berkeley.edu}
\affiliation{%
 \institution{UC Berkeley}
 \city{Berkeley}
 \state{California}
 \country{USA}}
 
\author{Madhav Marathe}
\author{Achla Marathe}
\author{Stephen Eubank}
\email{{mvm7hz,avm6bk,sge3qp}@virginia.edu}

\affiliation{%
  \institution{UVA Biocomplexity Institute}
  \streetaddress{P.O. Box 400298}
  \city{Charlottesville}
  \state{Virginia}
  \country{USA}
  \postcode{22904-4298}}

\author{Michael S. Bernstein}
\email{msb@stanford.edu}
\affiliation{%
  \institution{Stanford University}
  \city{Stanford}
  \state{California}
  \country{USA}
}

\renewcommand{\shortauthors}{Ritchie, et al.}

\begin{abstract}
  Beginning in April 2020, we gathered partial county-level data on non-pharmaceutical interventions (NPIs) implemented in response to the COVID-19 pandemic in the United States, using both volunteer and paid crowdsourcing. In this report, we document the data collection process and summarize our results, to increase the utility of our open data and inform the design of future rapid crowdsourcing data collection efforts. 
\end{abstract}



\maketitle

\vspace{0.45in}

\section{Introduction}

In early April 2020, we launched an effort to collect crowdsourced county-level data on policy interventions aimed at curbing the spread of COVID-19 in the United States.

This was prompted by the fact that, at the beginning of the COVID-19 pandemic, effective pharmaceutical interventions such as vaccines were not yet available. Public health authorities instead relied on a strategy of reducing transmission by limiting social contacts, using a class of policy interventions referred to as non-pharmaceutical interventions (NPIs). However, this policy response varied greatly in timing and scope. One reason for this is that in the United States, the authority for imposing public health restrictions is often held at the county level~\cite{future1988}. Though in almost all areas, state-wide ordinances were eventually put in place, this often occurred after county-level measures were implemented, and the exact nature of the restrictions often differed. As one concrete example, a judge in Harris County, Texas issued a stay-at-home order that took effect on March 24, 2020, while a similar state-wide order did not come into effect in Texas until April 2, 2020. 



Given the large number of U.S. counties and county equivalents (3,243 in 2020), this heterogeneity posed both a challenge for data collection and an opportunity for comparative causal analyses that could give estimates of the effectiveness of various policies.

To address this data collection challenge, we initially focused on volunteer-driven crowdsourcing of information on policies in different counties. Based on the distribution of responses, which had significant gaps in areas of the country with low population density, we then conducted paid crowdsourcing by employing workers on Amazon Mechanical Turk (AMT) to explore whether this would allow us to gather country-wide data.

We are sharing this report in order to document the process by which we collected our data, to facilitate analysis and re-use of our open dataset (\url{http://hci.st/socialdistancing-dataset}), and to share the challenges encountered during the process with researchers interested in developing tools for rapid crowdsourced data collection in response to developing public health crises. 

\section{Survey Design}
We began the project on March 23, 2020, less than two weeks after the WHO declared COVID-19 a pandemic on March 11. Given the need to collect county-level data on non-pharmaceutical interventions, volunteer crowdsourcing seemed like a promising option, as we expected people would already be familiar with the NPIs in place in their local area, or else be able to leverage their familiarity with local information sources to gather that information effectively. 

After considering a variety of alternative designs (including wiki-style interfaces and shared spreadsheets), we decided that a survey implemented using the Qualtrics platform was the best option for our project, since it would allow us to begin data collection quickly, without the development and extensive testing that a custom solution would have entailed. 

In addition to questions on local non-pharmaceutical interventions, we also asked respondents questions about their personal compliance with local NPIs and the estimated level of compliance in their immediate area. It was unclear in the early stage of the pandemics the degree to which populations would comply with restrictive NPIs, and whether this would vary geographically.  

If we already had three or more responses for a given county, we directed respondents directly to the compliance questions, since we anticipated that a larger sample size would be needed to obtain meaningful results. We conducted several rounds of data collection and updated the per-county counts of valid responses after each round was complete. As a consequence, counties that surpassed the three-response threshold during a round continued to receive more responses, and sometimes exceeded our requirements.

\subsection{Local NPI Questions}
We focused on the following types of policies, relying on the domain expertise of the authors with backgrounds in epidemiology and pandemic control.

\begin{itemize}
 \item Non-Essential Business Closures
 \item K-12 School Closures
 \item College / University Closures
 \item Religious Services Closures
 \item Lockdown or "Stay-at-home" Orders
\end{itemize}

Notably, we did not consider mask mandates or other policies related to mask-wearing, because at the initial stage of the pandemic when we were designing our survey, such policies had not been widely implemented. Guidance from the CDC discouraged widespread mask wearing until April 3rd, and the first state-wide mask mandate was issued on April 10th.

For each policy, we asked volunteers if the NPI had taken effect in their county, if it had ended at the time they completed the survey, and to provide a start date and end date if appropriate. We encouraged volunteers to add a source URL to allow verification of their answers, but did not require it, because we anticipated that this would decrease the number of responses we might receive.

We asked that volunteers follow a number of conventions when answering the survey --- for example, if there were multiple dates (e.g. some schools closed earlier than others) we asked volunteers to pick the earliest one they were aware of. For school closures, we asked that the recorded start date match the first day students were not in school (including weekends and spring break periods), rather than the first missed school day after the policy took effect. For religious service closures, we asked for the earliest date, including both voluntary and government-imposed closures. 

\subsection{Local Compliance Estimates}

If respondents indicated that they were reporting information on a county were they currently lived, we asked optional questions about their own social distancing practices and what they had observed around them. We included three questions that used a critical incident analysis strategy by asking participants about how recently they had left their home, how recently they had seen someone in their area violate social distancing guidelines (defined as not maintaining a six foot distance) and how recently they had personally come within six feet of someone outside of their household.



\section{Survey Distribution}

\begin{figure}
 \includegraphics[width=\columnwidth]{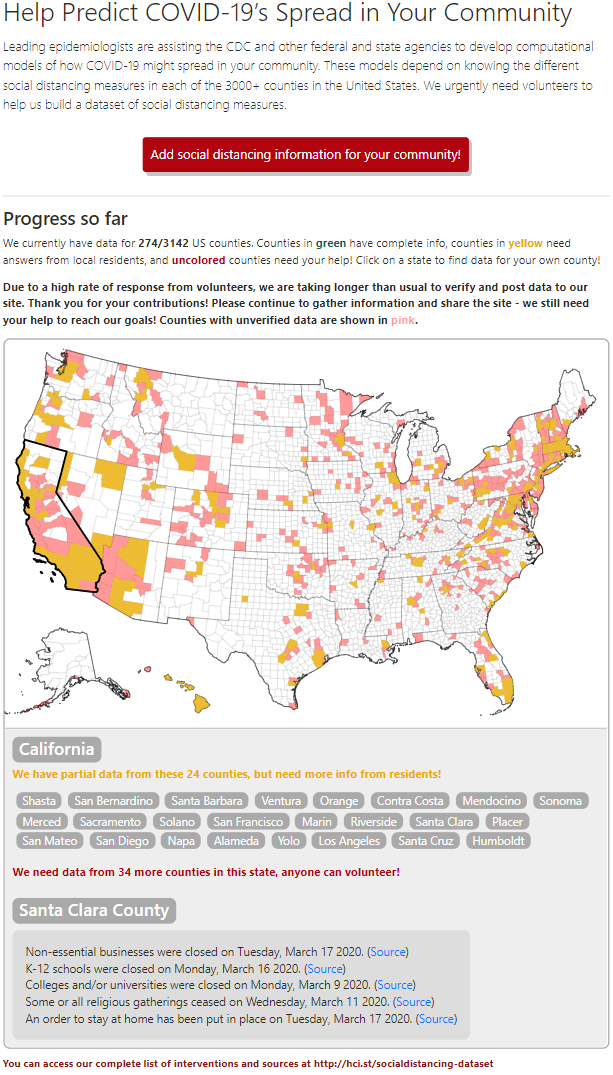}
 \caption{Call to action from \url{socialdistancing.stanford.com}, with visualization of in-progress data collection and the ability to view dates and source URLs for each county.}
 \label{fig:splash}
\end{figure}

We designed a landing page with a prominent call-to-action, and a visualization of current data collection progress. This interactive visualization also allowed visitors to the page to see the data we had collected (see Figure~\ref{fig:splash}). The visualization was updated intermittently with new data.

We first added data gathered by the development team while piloting the survey, and then distributed it to friends and family to collect initial responses. This was done in an attempt to avoid a "cold-start" problem, as we thought that volunteers would be more likely to contribute if they saw that some data had already been populated. Entries for 50 counties were completed before the public launch. 

On April 6th 2020, we launched the site publicly, and promoted it on various social media platforms. We also worked with the Stanford Engineering communications office to create a press-release describing the crowdsourcing effort, and then did another round of social media promotion after this launched on April 13th 2020.

\section{Volunteer Crowdsourcing Results}

\begin{figure}
 \includegraphics[width=\columnwidth]{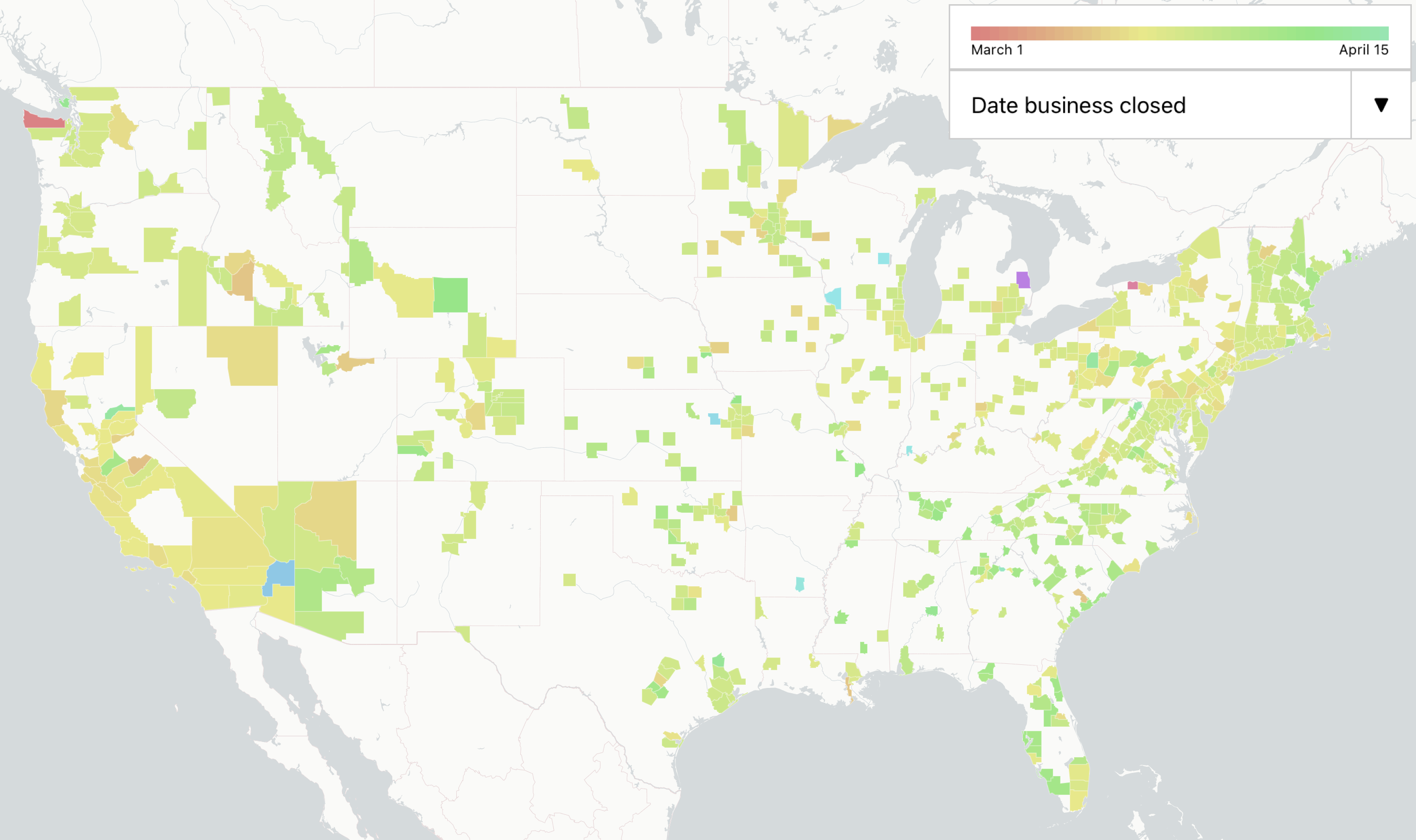}
 \caption{Reported Business Closure dates by county, for counties where we collected at least one report. Where multiple reports conflicted, we used the \emph{median} date. Note: because we asked about current restrictions at the time of survey completion, most of the dates were in the period of March and April 2020, when the majority of data was collected, with a small number of responses from months later. } 
 \label{fig:business-close-dates}
\end{figure}

We received 5,318 responses to our survey following the public launch. Our final crowdsourced dataset includes information for 415 US counties. This includes the initial responses that were populated by the team members and pilot participants prior to the public launch. Figure~\ref{fig:business-close-dates} shows the geographic distribution of information gathered for one NPI category (Business Closure dates).

The dataset consists mainly of information on policy start dates, with policy end dates missing for almost all responses, usually because the NPIs had not been lifted when the response was added or verified. We did not explicitly ask verifiers to seek out policy end date information that was not present in the volunteer responses.

There was a high rate of error in the volunteer responses. Common error types included using information on the wrong type of policy (e.g. using school closure information in place of a lockdown order), using the publication date of an order or news article instead of the date that the policy took place, not finding the earliest order that applied to the county, not following conventions to account for weekends and spring break periods, and not distinguishing between closure of non-essential business closures and narrower closure orders that only affected high-risk businesses. 

Some but not all errors were caught by the verification process, and caution should still be taken regarding all of these potential error sources when using the data. 

Initially, responses were verified by the Stanford team members that designed the survey. The verification workflow used by the Stanford team members did not follow an explicit protocol. Later, employees at the University of Virginia Biocomplexity Institute validated dates and sources using a workflow developed by the survey team (see appendix for complete workflow). The counties verified by the UVA team members had received at least three partial responses, or 1-2 responses which contained source URLs for each completed question. 

We include a source URL for each start or end date in the dataset whenever possible, but some entries do not have URLs, either due to oversights in the verification process, or rare cases when several volunteer respondents agreed on the date but the verifier was unable to locate a source.

\section{Paid Crowdsourcing}
While we did receive a large number of responses, we concluded that we were unlikely to gather complete country-wide data through volunteer responses alone. We obtained a grant from Amazon Mechanical Turk to fund further data collection by paid crowd workers. Workers were compensated at a rate of \$15.00 per hour using the Fair Work script~\cite{whiting2019fair}.

We chose 3 states for the initial paid crowdsourcing: Georgia, California and Massachusetts, comprising 184 counties. These states were chosen because we had access to data on NPIs for some or all of the counties in those states. This allowed us to assess the level of error in the collected responses. 

For California and Massachusetts, we used the volunteer crowdsourced data as a point of comparison. For Georgia, we relied on county-level data gathered by the Center for the Ecology of Infectious Diseases Coronavirus Working Group at UGA~\cite{UGA}. We asked workers to find the earliest occurrence of each type of NPI. 

In addition to using paid crowdsourcing to scale up data collection, we were interested in whether we could reduce the error level in the resulting data without relying on a dedicated team of volunteers for verification, as we had done in the prior phase of the project. We considered two strategies for achieving this goal - producing an estimated start date for each NPI category by aggregating crowdsourcing responses following a "wisdom of the crowds" approach~\cite{surowiecki2005wisdom}, and  creating additional AMT workflows for explicit verification of policy start dates. 

We performed several experiments to explore the accuracy of our crowdsourced estimates and how to aggregate the dates we received. We initially restricted each task to workers residing in the state where we were gathering policy information. We then tested the effect of allowing the most reliable workers (based on this initial data collection) to collect data for counties in other states.

We developed two AMT workflows for verification. One was intended to be faster to complete, asking workers to check whether the date could be found in any of the provided sources (lightweight method), while the other (checklist method) guided them through a series of questions based off an analysis of common errors in a subset of questions.


\section{Paid Crowdsourcing Results}

\subsection{Accuracy by number of responses and aggregator choice}
We considered two important dimensions in analyzing the paid crowdsourcing data: the number of worker responses per county and the choice of aggregator. The analysis based on the first round of data collection informed design choices for subsequent rounds of data collection, as it enabled us to estimate the number of responses needed per county to achieve an acceptable error rate.



\begin{figure}
 \includegraphics[width=\columnwidth]{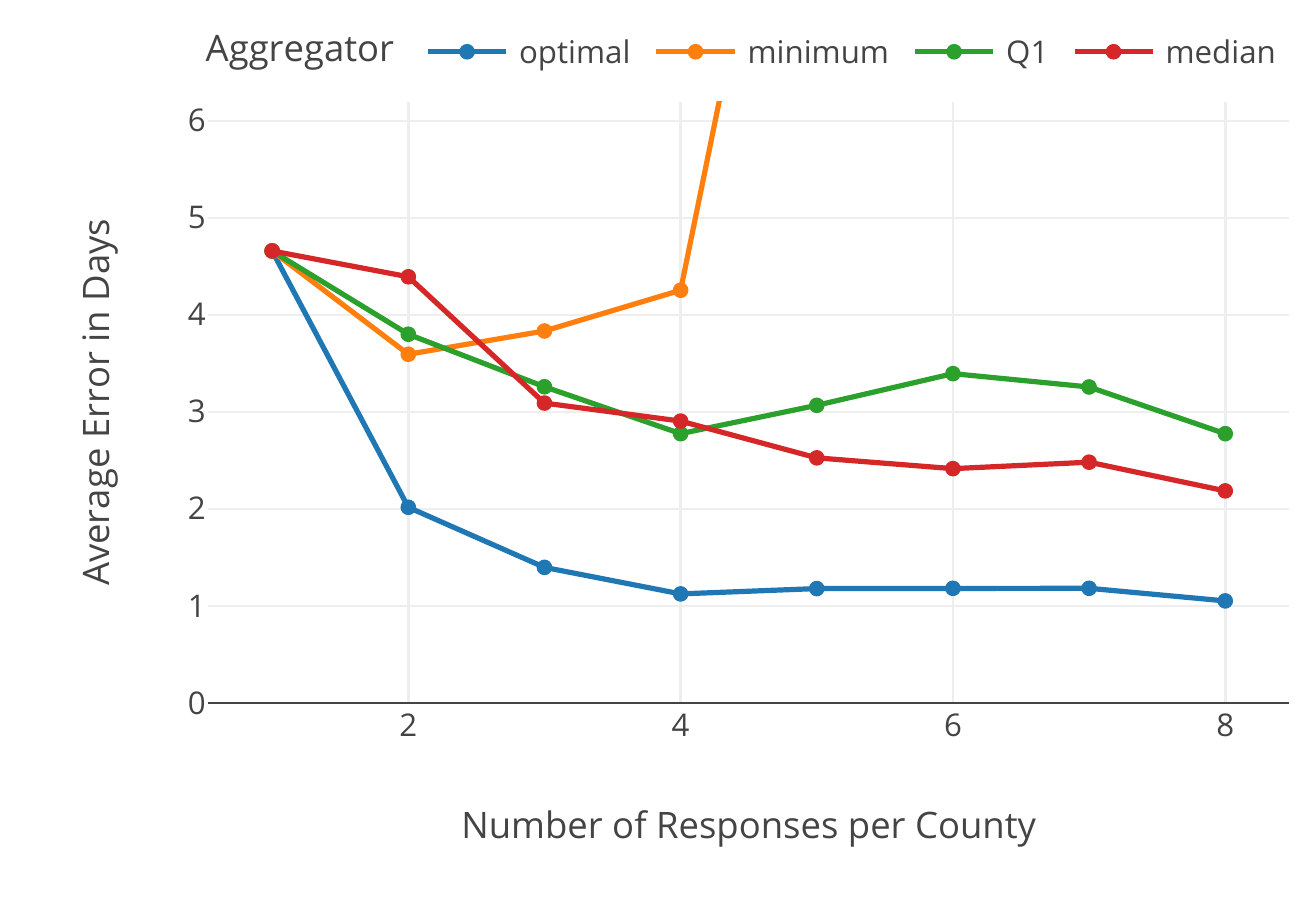}
 \caption{Performance of various aggregators as a function of the number of responses per county.}
 \label{fig:heterogeneity_by_agg_and_num_responses}
\end{figure}

In Georgia, Massachusetts, and California, we collected up to eight responses per county and computed the average error in days using four types of aggregators:
\begin{itemize}
    \item The \emph{optimal} aggregator: selecting the submitted date that is closest to the verified true date. This requires the availability of human-verified dates, and so is shown only for comparison purposes, as it allows for error decomposition into the component that arises from the aggregator choice and the component that comes from not having correct responses in the data set.
    \item The \emph{minimum}, \emph{Q1}, and \emph{median} aggregators: taking the earliest date, the 25th percentile, and the median date from the responses recorded in that county, respectively. We considered the \emph{minimum} and \emph{Q1} aggregators because of the assumption that if one or more workers found an earlier date, it was more likely to be a county-specific policy rather than a state-level policy. 
\end{itemize}
To estimate the average error that would have been observed with fewer responses, we randomly removed workers' responses from our data before aggregation. The resulting curves in Figure~\ref{fig:heterogeneity_by_agg_and_num_responses} show the average error in days for each aggregator, as a function of the number of responses per county. 


\begin{figure*}[h]
 \includegraphics[width=\textwidth]{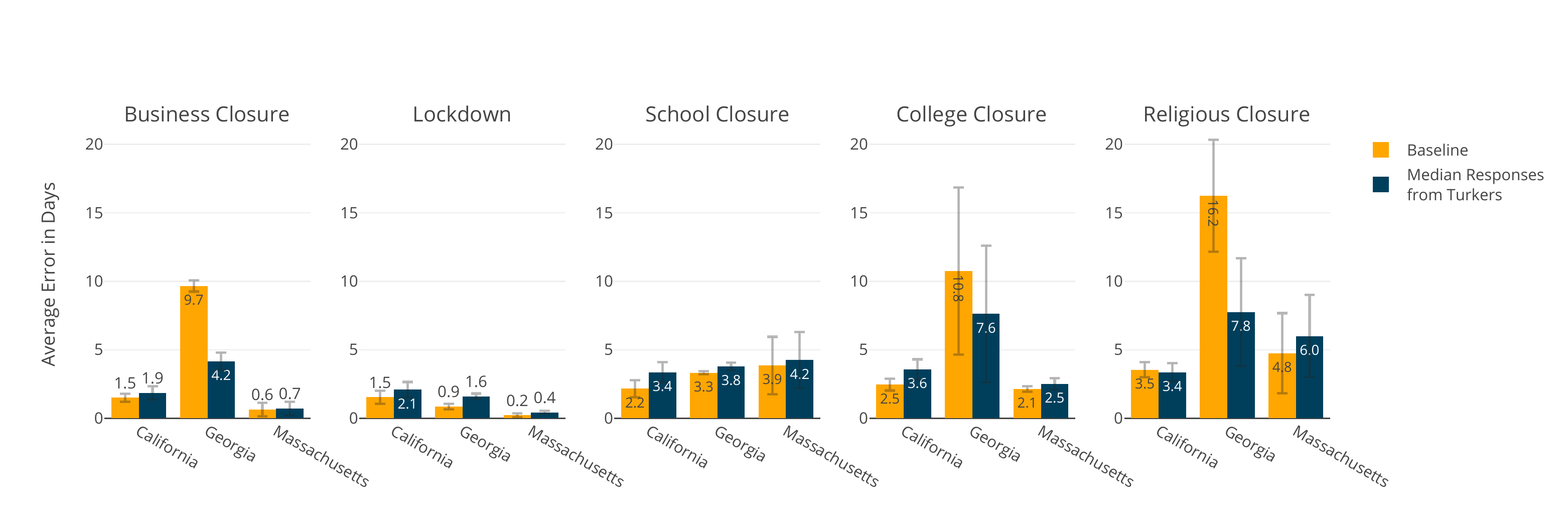}
 \caption{Comparison of average error between baseline and \emph{median} responses from crowd workers for various combinations of state and NPI type. The baseline for Business Closure, Lockdown, and Religious Closure questions is the statewide lockdown date, whereas the baseline for School Closure and College Closure questions is the statewide school closure date.}
 \label{fig:heterogeneity_by_state_and_question_type}
\end{figure*}

The optimal aggregator curve indicates that on average, the error in the best case based on eight responses per county could be brought down to $\pm$1 day using verification, using only the information contained in the original worker submissions.
The error using the median of submitted dates as an estimator decreases consistently until we get to eight responses per county ($\pm$2 days). However, we only observed a marginal benefit to collecting more than three responses.
The minimum estimator is not robust and the error tends to go up quickly as the number of responses increases.

Based on this analysis, we decided to collect three responses per county and used the median aggregator in subsequent rounds of data collection.

\subsection{Accuracy heterogeneity by state and question type}

Using our verified data for 47 counties (as well as School Closure, Lockdown and Business Closure data for Georgia from UGA), we quantified the level of error present in the crowdsourced policy dates across states and question types. The baseline for Business Closure, Lockdown, and Religious Closure questions is the statewide lockdown date, whereas the baseline for School Closure and College Closure questions is the statewide school closure date.

Figure~\ref{fig:heterogeneity_by_state_and_question_type} shows that data aggregated from AMT significantly reduced the error for Georgia for the Business Closure question and the Religious Closure question.

 Without correction for multiple comparisons, we cannot show that our Mechanical Turk workflow is able to significantly lower the error rate in general, but this does provide evidence that it might be useful for some combinations of state and policy type \cite{whentocorrect}. California and Massachusetts displayed relatively low variance in initial NPI roll-out dates, while Georgia had much higher variance. Thus, we can form a post-hoc hypothesis that county-level data from paid crowdsourcing might outperform the state-wide policy information baseline for certain states with high heterogeneity at the county-level. Other states with high levels of variation include Missouri, Nebraska, Tennessee, and Pennsylvania~\cite{covid-vis}.

\subsection{Prioritizing the most reliable turkers}
We noticed that the error level varied widely depending on which AMT worker completed the task. We performed an experiment to see if this could be used to improve the quality of our results. Sixty-four workers with low average error in previous tasks (< 3 days) were given a custom qualification. This pool of workers provided four additional responses for each county in Massachusetts, and we compared their responses to those provided by in-state workers without such qualification. However, we did not find significant differences when comparing error rates among the two groups.

Interestingly, for this round of data collection, taking the \emph{minimum} of the response dates resulted in lower average error than the \emph{median} aggregator (contrary to what we observed in the initial round of paid crowdsourcing). In addition to the changes to worker qualifications, one factor that may have contributed to this shift is that more time had elapsed since the NPIs of interest had been implemented. The bias in the \emph{minimum} aggregator might have compensated for the increased likelihood of workers reporting more recent interventions instead of the first intervention to occur in a county.    


\subsection{Error analysis}
We analyzed the specific errors found in the a subset of the crowdsourced data to inform future survey designs, and found that the following were among the most common classes of error:
\begin{itemize}
    \item Using information on more recent NPIs
    \item Not using reliable data sources such as official orders
    \item Data input errors (e.g. reporting the month incorrectly)
    \item Using the article publication date rather than the date the policy took effect.
\end{itemize}
We subsequently revised the data collection instructions to point out these common mistakes and encourage workers to avoid them. A large portion of the error seems to be caused by outlier dates (due to data input errors or using more recent NPIs), which could be prevented by enforcing a pre-defined range for the policy start dates that workers can report, either during the data collection process or a verification process.

\subsection{Comparing lightweight and checklist-based verification workflows}
We developed two AMT workflows for verification. One was intended to be faster to complete, asking workers to check whether the date could be found in any of the provided sources (the \emph{lightweight method}), while the other (the \emph{checklist method}, or "heavyweight" method) guided them through a series of questions based on the analysis of common errors reported in the previous section.

We found that while we could greatly reduce the error in the data through verification, the number of verifiers required to meaningfully reduce the average error (five) was large enough for this process to be cost-prohibitive. The unit cost to verify a single question was \$3.68, meaning that each county would cost \$18.40 per verifier. Because of this, we decided aggregate the initial responses rather than relying on an explicit verification workflow.

We did not observe large differences between the error rates in the results obtained via the lightweight and checklist methods. 




\section{Follow-Up Data Collection}

We gathered additional NPI data through paid crowdsourcing in February 2021, following the same methods used in the previous rounds of data collection. We prioritized counties with high \emph{per capita} rates of COVID infection. Due to the much larger time interval that had elapsed since the introduction of county-level NPIs early in the pandemic, we expect that the error rate in this data is higher than what was observed in earlier rounds of data collection.


\section{Discussion and Future Work}

How to employ untrained volunteers or crowd workers to collect complex policy information, while ensuring a reasonable level of data quality, remains an important area for future investigation. 

In general, we found that the verification process for data collected in this way was time consuming, often requiring more time than the initial data collection. In particular, the most challenging aspect of the process was that verifiers needed to search for relevant dates in the provided source documents (or find a missing source if no URL was provided), and then parse the language in that document to establish its correctness, which might not be trivial. Consider the sentence "The directive will take effect upon close of business Friday", which requires the verifier to consider the publication date of the article. 
Giving respondents tools for annotating the specific passages in a source document that support claims about NPI start and end dates might be useful for improving the efficiency of the verification process. 

We also explored soliciting a large number of responses through paid crowdsourcing and aggregating the resulting policy start date information without an explicit verification process. However, besides requiring much greater amounts of crowdsourced data, it is unclear how to validate whether the error levels from aggregated estimates are acceptable, especially if finding high-quality ground truth data is difficult, as was the case for data on NPI measures at the beginning of the pandemic. Our results also suggest that the relative performance of different aggregation methods might vary depending on the time interval that has elapsed between the policy event of interest and data collection. 

Another possible alternative is to increase the initial data quality level by relying on trained volunteers or crowd workers to gather data, rather than allowing a large number of novices to contribute policy information. This strategy resulted in high data quality in another crowdsourced county-level NPI data collection effort~\cite{hms-dataset}. 

Even if the error rate was reduced through these or similar measures, a more complicated database schema would be necessary to capture the full complexity of the NPI rollout. Our assumption that for each NPI there would be a single unambiguous start and end date was often inaccurate because of variation below the county level. Municipalities often imposed their own restrictions. K-12 school closures often occurred at the school board level, which do not map precisely onto counties or municipalities. College closures were decided by individual schools, and some counties contained multiple colleges, or none. 

We attempted to add conventions to our volunteer and crowd worker instructions and verification instructions to resolve these and other related difficulties. However, adding clarifications and caveats to these workflows made them more cognitively demanding, and this type of guidance was sometimes ignored. This might suggest a need for different workflows for each type of policy intervention that take into account the different relevant geographic entities, rather than trying to gather all policy information at the county level using a common workflow. 

Another limitation of the dataset is that over time the source URLs may become outdated due to "link rot" and the information used by respondent or verifier may no longer be accessible \cite{taylor2000linkrot}. To avoid this in future data collection efforts, automated or manual retrieval and storage of the source documents themselves, rather than URLs, would be ideal. 

Similarly, the data collection task becomes more difficult as time progresses, since volunteers and crowd workers may have a harder time accessing relevant sources, and more irrelevant material (for example, news stories about subsequent policy interventions) will be present. This indicates a need for rapid response when collecting this type of policy information.


A final challenge we encountered, that is likely to recur in future crowdsourcing efforts, was the difficulty of reaching remote and less populous counties. We relied mainly on promotion through social media, starting with our own personal and professional networks. This resulted in more responses in California and Virginia, where most of our team was based. In comparison, Figure 1 shows that large areas of the central US received no responses. This could have been ameliorated through other forms of volunteer recruitment, such as targeted advertising. However even with such strategies, reaching rural areas will remain challenging, since there are simply less potential volunteers, and finding information is complicated by lower levels of news media coverage. Our response to this issue was to explore paid crowdsourcing to fill gaps in volunteer recruitment, but crowd workers will be less familiar with local policies and information sources than local volunteers. Leveraging crowdsourcing to serve these areas of low population density is a potential research area for the emerging subdiscipline of Rural HCI~\cite{10.1145/3359298}.

\section{Other Data Collection Efforts}
In parallel to our efforts, other parties collected similar county-level
NPI datasets, some of them using crowdsourcing methodologies. Ebrahim et. al. led an effort using trained volunteers to collect information on NPIs in 1,320 US counties~\cite{hms-dataset}. Keystone Strategy led an NPI collection effort focused on the most populous US counties~\cite{Keystone}. The COVIDVIS project also collected a number of NPI dates~\cite{covid-vis}. The Center for the Ecology of Infectious Diseases Coronavirus Working Group gathered detailed county-level NPI information for Georgia~\cite{UGA}. The HIT-COVID Collaboration, a large (200+ collaborator) volunteer team, created the dataset described in Zheng et al.  which contains over 10,000 records, including the majority of measures implemented in most African countries.    \cite{zheng2020hit} Complexity Science Hub Vienna gathered hierarchically clustered information on NPIs in 57 countries, over 6,000 records, including partial information on regional measures in the United States \cite{desvars2020structured}. This data was used to compare the effectiveness of NPI measures worldwide \cite{haug2020ranking}.



Cross-referencing and synthesizing these datasets may allow for detection of errors and for coverage of a larger number of counties across longer timescales.






\begin{acks}
This data collection effort would not have been possible without the help of the following individuals who assisted with data verification on top of their normal work responsibilities: Golda Barrow, James Walke, Benjamin Selwyn, Jilian Draughon, Erin Raymond, Lona Doyle, and Lori Conerly.

We also wish to acknowledge Jackie Yang, Tianshi Li, Mitchell Gordon, Amy Zhang, Catherine Mullings, Rajan Vaish, Golrokh Emami, Dana\"{e} Metaxa and Mandy Wilson for their contributions to the project, as well as Mallory Harris and Madalsa Singh for their comments on drafts of this report. 

Finally, we wish to acknowledge a grant of credits from Amazon Mechanical Turk that enabled us to perform paid data collection.
\end{acks}

\bibliographystyle{ACM-Reference-Format}
\bibliography{references}


\begin{thebibliography}{13}


\ifx \showCODEN    \undefined \def \showCODEN     #1{\unskip}     \fi
\ifx \showDOI      \undefined \def \showDOI       #1{#1}\fi
\ifx \showISBNx    \undefined \def \showISBNx     #1{\unskip}     \fi
\ifx \showISBNxiii \undefined \def \showISBNxiii  #1{\unskip}     \fi
\ifx \showISSN     \undefined \def \showISSN      #1{\unskip}     \fi
\ifx \showLCCN     \undefined \def \showLCCN      #1{\unskip}     \fi
\ifx \shownote     \undefined \def \shownote      #1{#1}          \fi
\ifx \showarticletitle \undefined \def \showarticletitle #1{#1}   \fi
\ifx \showURL      \undefined \def \showURL       {\relax}        \fi
\providecommand\bibfield[2]{#2}
\providecommand\bibinfo[2]{#2}
\providecommand\natexlab[1]{#1}
\providecommand\showeprint[2][]{arXiv:#2}

\bibitem[\protect\citeauthoryear{Desvars-Larrive, Dervic, Haug,
  Niederkrotenthaler, Chen, Di~Natale, Lasser, Gliga, Roux, Sorger,
  et~al\mbox{.}}{Desvars-Larrive et~al\mbox{.}}{2020}]%
        {desvars2020structured}
\bibfield{author}{\bibinfo{person}{Amelie Desvars-Larrive},
  \bibinfo{person}{Elma Dervic}, \bibinfo{person}{Nils Haug},
  \bibinfo{person}{Thomas Niederkrotenthaler}, \bibinfo{person}{Jiaying Chen},
  \bibinfo{person}{Anna Di~Natale}, \bibinfo{person}{Jana Lasser},
  \bibinfo{person}{Diana~S Gliga}, \bibinfo{person}{Alexandra Roux},
  \bibinfo{person}{Johannes Sorger}, {et~al\mbox{.}}}
  \bibinfo{year}{2020}\natexlab{}.
\newblock \showarticletitle{A structured open dataset of government
  interventions in response to COVID-19}.
\newblock \bibinfo{journal}{\emph{Scientific Data}} \bibinfo{volume}{7},
  \bibinfo{number}{1} (\bibinfo{year}{2020}), \bibinfo{pages}{1--9}.
\newblock


\bibitem[\protect\citeauthoryear{Ebrahim, Ashworth, Noah, Kadambi, Toumi, and
  Chhatwal}{Ebrahim et~al\mbox{.}}{2020}]%
        {hms-dataset}
\bibfield{author}{\bibinfo{person}{Senan Ebrahim}, \bibinfo{person}{Henry
  Ashworth}, \bibinfo{person}{Cray Noah}, \bibinfo{person}{Adesh Kadambi},
  \bibinfo{person}{Asmae Toumi}, {and} \bibinfo{person}{Jagpreet Chhatwal}.}
  \bibinfo{year}{2020}\natexlab{}.
\newblock \showarticletitle{Reduction of COVID-19 Incidence and
  Nonpharmacologic Interventions: Analysis Using a US County--Level Policy Data
  Set}.
\newblock \bibinfo{journal}{\emph{J Med Internet Res}} \bibinfo{volume}{22},
  \bibinfo{number}{12} (\bibinfo{date}{21 Dec} \bibinfo{year}{2020}),
  \bibinfo{pages}{e24614}.
\newblock
\showISSN{1438-8871}
\urldef\tempurl%
\url{https://doi.org/10.2196/24614}
\showDOI{\tempurl}


\bibitem[\protect\citeauthoryear{for the Ecology of Infectious Diseases
  Coronavirus Working Group at the University~of Georgia}{for the Ecology of
  Infectious Diseases Coronavirus Working Group at the University~of
  Georgia}{2020}]%
        {UGA}
\bibfield{author}{\bibinfo{person}{Center for the Ecology of Infectious
  Diseases Coronavirus Working Group at the University~of Georgia}.}
  \bibinfo{year}{2020}\natexlab{}.
\newblock \bibinfo{title}{COVID-19 Data}.
\newblock
  \bibinfo{howpublished}{\url{https://github.com/CEIDatUGA/COVID-19-DATA}}.
\newblock


\bibitem[\protect\citeauthoryear{Hardy, Wyche, and Veinot}{Hardy
  et~al\mbox{.}}{2019}]%
        {10.1145/3359298}
\bibfield{author}{\bibinfo{person}{Jean Hardy}, \bibinfo{person}{Susan Wyche},
  {and} \bibinfo{person}{Tiffany Veinot}.} \bibinfo{year}{2019}\natexlab{}.
\newblock \showarticletitle{Rural HCI Research: Definitions, Distinctions,
  Methods, and Opportunities}.
\newblock  \bibinfo{volume}{3}, \bibinfo{number}{CSCW}, Article
  \bibinfo{articleno}{196} (\bibinfo{date}{nov} \bibinfo{year}{2019}),
  \bibinfo{numpages}{33}~pages.
\newblock
\urldef\tempurl%
\url{https://doi.org/10.1145/3359298}
\showDOI{\tempurl}


\bibitem[\protect\citeauthoryear{Haug, Geyrhofer, Londei, Dervic,
  Desvars-Larrive, Loreto, Pinior, Thurner, and Klimek}{Haug
  et~al\mbox{.}}{2020}]%
        {haug2020ranking}
\bibfield{author}{\bibinfo{person}{Nils Haug}, \bibinfo{person}{Lukas
  Geyrhofer}, \bibinfo{person}{Alessandro Londei}, \bibinfo{person}{Elma
  Dervic}, \bibinfo{person}{Am{\'e}lie Desvars-Larrive},
  \bibinfo{person}{Vittorio Loreto}, \bibinfo{person}{Beate Pinior},
  \bibinfo{person}{Stefan Thurner}, {and} \bibinfo{person}{Peter Klimek}.}
  \bibinfo{year}{2020}\natexlab{}.
\newblock \showarticletitle{Ranking the effectiveness of worldwide COVID-19
  government interventions}.
\newblock \bibinfo{journal}{\emph{Nature human behaviour}} \bibinfo{volume}{4},
  \bibinfo{number}{12} (\bibinfo{year}{2020}), \bibinfo{pages}{1303--1312}.
\newblock


\bibitem[\protect\citeauthoryear{of~Medicine (US) Committee for the Study of
  the Future~of Public~Health}{of~Medicine (US) Committee for the Study of the
  Future~of Public~Health}{1988}]%
        {future1988}
\bibfield{author}{\bibinfo{person}{Institute of~Medicine (US) Committee for the
  Study of the Future~of Public~Health}.} \bibinfo{year}{1988}\natexlab{}.
\newblock \bibinfo{booktitle}{\emph{The Future of Public Health}}.
\newblock \bibinfo{publisher}{National Academies Press (US)},
  \bibinfo{address}{Washington, D.C.}
\newblock
\urldef\tempurl%
\url{https://www.ncbi.nlm.nih.gov/books/NBK218212/}
\showURL{%
\tempurl}


\bibitem[\protect\citeauthoryear{Rubin}{Rubin}{2021}]%
        {whentocorrect}
\bibfield{author}{\bibinfo{person}{Mark Rubin}.}
  \bibinfo{year}{2021}\natexlab{}.
\newblock \showarticletitle{When to adjust alpha during multiple testing: A
  consideration of disjunction, conjunction, and individual testing}.
\newblock \bibinfo{journal}{\emph{Synthese}} (\bibinfo{year}{2021}),
  \bibinfo{pages}{1--32}.
\newblock


\bibitem[\protect\citeauthoryear{Strategy}{Strategy}{2020}]%
        {Keystone}
\bibfield{author}{\bibinfo{person}{Keystone Strategy}.}
  \bibinfo{year}{2020}\natexlab{}.
\newblock \bibinfo{title}{COVID-19 Intervention Data}.
\newblock
  \bibinfo{howpublished}{\url{https://github.com/Keystone-Strategy/covid19-intervention-data}}.
\newblock


\bibitem[\protect\citeauthoryear{Surowiecki}{Surowiecki}{2005}]%
        {surowiecki2005wisdom}
\bibfield{author}{\bibinfo{person}{James Surowiecki}.}
  \bibinfo{year}{2005}\natexlab{}.
\newblock \bibinfo{booktitle}{\emph{The wisdom of crowds}}.
\newblock \bibinfo{publisher}{Anchor}.
\newblock


\bibitem[\protect\citeauthoryear{Taylor and Hudson}{Taylor and Hudson}{2000}]%
        {taylor2000linkrot}
\bibfield{author}{\bibinfo{person}{Mary~K Taylor} {and} \bibinfo{person}{Diane
  Hudson}.} \bibinfo{year}{2000}\natexlab{}.
\newblock \showarticletitle{" Linkrot" and the usefulness of Web site
  bibliographies}.
\newblock \bibinfo{journal}{\emph{Reference \& User Services Quarterly}}
  (\bibinfo{year}{2000}), \bibinfo{pages}{273--277}.
\newblock


\bibitem[\protect\citeauthoryear{Team}{Team}{2020}]%
        {covid-vis}
\bibfield{author}{\bibinfo{person}{Covidvis Team}.}
  \bibinfo{year}{2020}\natexlab{}.
\newblock \bibinfo{title}{COVID-19-Vis}.
\newblock
  \bibinfo{howpublished}{\url{https://github.com/covidvis/covid19-vis}}.
\newblock


\bibitem[\protect\citeauthoryear{Whiting, Hugh, and Bernstein}{Whiting
  et~al\mbox{.}}{2019}]%
        {whiting2019fair}
\bibfield{author}{\bibinfo{person}{Mark~E Whiting}, \bibinfo{person}{Grant
  Hugh}, {and} \bibinfo{person}{Michael~S Bernstein}.}
  \bibinfo{year}{2019}\natexlab{}.
\newblock \showarticletitle{Fair work: Crowd work minimum wage with one line of
  code}. In \bibinfo{booktitle}{\emph{Proceedings of the AAAI Conference on
  Human Computation and Crowdsourcing}}, Vol.~\bibinfo{volume}{7}.
  \bibinfo{pages}{197--206}.
\newblock


\bibitem[\protect\citeauthoryear{Zheng, Jones, Leavitt, Ung, Labrique, Peters,
  Lee, and Azman}{Zheng et~al\mbox{.}}{2020}]%
        {zheng2020hit}
\bibfield{author}{\bibinfo{person}{Qulu Zheng}, \bibinfo{person}{Forrest~K
  Jones}, \bibinfo{person}{Sarah~V Leavitt}, \bibinfo{person}{Lawson Ung},
  \bibinfo{person}{Alain~B Labrique}, \bibinfo{person}{David~H Peters},
  \bibinfo{person}{Elizabeth~C Lee}, {and} \bibinfo{person}{Andrew~S Azman}.}
  \bibinfo{year}{2020}\natexlab{}.
\newblock \showarticletitle{HIT-COVID, a global database tracking public health
  interventions to COVID-19}.
\newblock \bibinfo{journal}{\emph{Scientific data}} \bibinfo{volume}{7},
  \bibinfo{number}{1} (\bibinfo{year}{2020}), \bibinfo{pages}{1--8}.
\newblock


\end{thebibliography}

\clearpage


\section*{Supplementary material}

# socialdistancing.stanford.edu - verification instructions v1.0

Start with the state name and county name that you are verifying.

Search the Qualtrics output spreadsheet using "CTRL-F" to find the data on the county

Create a new row on the dataset spreadsheet ([http://hci.st/socialdistancing-dataset](http://hci.st/socialdistancing-dataset)) using "Right-click > Insert row". Keep the dataset spreadsheet sorted in alphabetical order by state and county.

There are six questions to verify.
- Have non-essential businesses been ordered to close?
- Have local K-12 schools closed?
- Have local colleges and universities closed?
- Have any in-person religious gatherings been canceled?.
- Have officials ordered residents to stay at home except for essential activities?

You should also check the comments field for useful information, but you can ignore all other questions. Link to full question list

For each of the six questions, there are six subfields. For example, for the non-essential business closure question:

- business_closed - Yes/No, have non-essential businesses in this community closed
- business_closed_date - Date (YYYY-MM-DD) - the date businesses closed
- business_closed_url - a source URL to verify this information
- business_open - Yes/No, have non-essential businesses in this community reopened
- business_open_date - Date (YYYY-MM-DD) - the date businesses reopened
- business_open_url - a source URL to verify reopening information

If business_closed is set to Yes, you should populate the business_closed_date and business_closed_url subfields (the same applies to business_open, business_open_date and business_open_url)

The other 5 questions have identical field structures (e.g. school_closed, college_closed, lockdown_closed, etc.).

The fields are the same in the Qualtrics output spreadsheet and the [http://hci.st/socialdistancing-dataset](http://hci.st/socialdistancing-dataset) spreadsheet, so you can copy information directly to the dataset spreadsheet once you have verified it.

If volunteers provided URLs, you should open that webpage and confirm the restriction and date information. Then add those three fields to the spreadsheet. Try to check every URL. You may have to adjust the date slightly (see Clarifications, below).

If volunteers do not provide URLs, but they generally agree that a restriction is in place, you need to find the missing date and source information manually.

Finding missing information

If http://hci.st/socialdistancing-dataset has information on other counties in the same state, and it agrees with the dates given by participants, check if the sources are consistent with a state-wide restriction. However, you should still look for a local news source or government posting that is county-specific if possible.

You can check this list of state executive orders, some of which apply to specific counties: https://web.csg.org/covid19/executive-orders/

You can use these Google queries to check each question:

**Non-essential business closure**
`[state name] [county name] businesses closed"`

**K-12 school closure**
`[state name] [county name] schools closed`

**College / university closure**
`[state name] [county name] college closure`
`[state name] [county name] university closure`

**Religious gathering closure**
`[state name] [county name] church online`

**Stay-at-home / shelter-in-place / lockdown**
`[state name] [county name] stay at home`
`[state name] [county name] shelter in place`

If the volunteer only provides a top-level URL (e.g. `cnn.com`), add that to the end of the search string.

Twitter is a good place to find university closure information. Facebook and Youtube are good places to find when religious services moved online.

If that doesn't find source information, try a few other queries. If you still can't find a source,

1. If there are several crowdsourced answers with the same date, use that date and leave the *_url field blank to indicate a missing source.
2. Find a source for the state-level restriction that applies (if any) and use the corresponding date.

If you can't find information about religious gathering cancellations, check if there is a gathering size limitation (e.g. max 10 people can gather in the same location). You can assume that this caused religious gathering closures unless there is evidence to the contrary (from your google search or from participant responses).

Clarifications

Religious gatherings might be canceled because the organization chose to cancel, or because officials required them to be canceled.

For school closures, college closures or religious gathering closures, we record the first date we can find for that county.

If multiple business closure or lockdown restrictions are in place in a county (for example, if there are two different towns that have different policies), choose the most restrictive policy you are aware of for the corresponding questions.

We record the date a restriction the order took effect (e.g., first day people were ordered to stay home, the first canceled Sunday church service), not the date the restriction was announced.

For school and university closures, we record the first day students were not in school
- If the first day schools were closed was a Monday, we put the Saturday of the previous week as the school closure date, since we assume students are not in class on weekends
- If the cancellation was announced while students were on Spring Break (often the case for both K-12 schools and universities), pick the first day of Spring Break. Search for [college name] spring break dates or [school board] spring break dates to find this information.

If an order only closes only a specific subset of businesses (e.g. bars, restaurants and movie theaters), we do not consider this as a non-essential business closure. We only record orders that close all businesses except a subset of essential businesses.












\end{document}